# A Systematic Literature Review on Technology Acceptance Research on Augmented Reality in the Field of Training and Education

Stefan Graser
CAEBUS Center for Advanced E-Business Studies
RheinMain University of Applied Science
Wiesbaden, Germany
e-mail: stefan.graser@hs-rm.de

Stephan Böhm
CAEBUS Center for Advanced E-Business Studies
RheinMain University of Applied Science
Wiesbaden, Germany
e-mail: stephan.boehm@hs-rm.de

*Abstract*—Augmented Reality (AR) is an emerging technology that ranks among the top innovations in interactive media. With the emergence of new technologies, the question about the factors influencing user acceptance arises. Many research models on the user acceptance of technologies were developed and extended to answer this question in the last decades. This research paper provides an overview of the current state in the scientific literature on user acceptance factors of AR in training and education. We conducted a systematic literature review, identifying 45 scientific papers on technology acceptance of augmented reality. Twenty-two papers refer more specifically to the field of training and education. Overall, 33 different technology acceptance models and 34 acceptance variables were identified. Based on the results, there is a great potential for further research.

## I. INTRODUCTION

AR is one of the top emerging technology innovations in interactive media. Over the past years, AR has found its way into different new application fields. One application area is education and training. AR can connect the digital and physical domains. Based on the technology, scenarios can be built to allow users to interact simultaneously in the real and virtual worlds. When applied to training and education, users can learn more effectively through new ways and methods using AR. Hence, education and training are one of the application fields where AR is applied most [1].

This research paper is about the status quo of the technology acceptance research on augmented reality as technology innovation in training and education. The main objective is to determine the state of research in the field mentioned above. The paper is structured as follows: Firstly, the theoretical background is described in Section II. For this, the terms AR and the theoretical foundation of technology acceptance, as well as is most common research models, are briefly described. Secondly, a systematic literature review was conducted. The scope and approach of this analysis are subject to Section III. The results of the literature review, including a more detailed discussion of the revealed models and variables, are presented in Section IV. The paper concludes with a summary of the findings and an outlook on further research in Section V.

## II. THEORETICAL FOUNDATION

### A. Augmented Reality

In the following, the theoretical background of augmented reality is explained. The beginnings of the development of AR solutions go back to the 1960s [2]. In scientific literature, Azuma [3] established a definition of AR: *"Augmented Reality (AR) is a variation of Virtual Environments (VE), or Virtual Reality [...], AR allows the user to see the real world, with virtual objects superimposed upon or composited with the real world."* So AR is an extension of reality, not a replacement for it. Furthermore, three key characteristics of AR are defined: Firstly, AR combines reality and virtuality. Secondly, AR is an interactive application in real-time. Thirdly, the content is shown in 3D.

AR is related to Virtual Reality (VR). Milgram et al. [4] imply a representation of this relation. This so-called Reality-Virtuality (RV) Continuum shown in Figure 1 combines AR/VR and introduces Mixed Reality (MR) in between. AR is next to the real environment, Augmented Virtuality (AV) to the virtual environment.

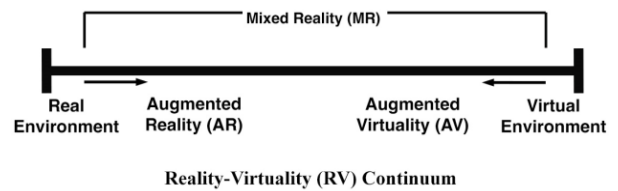

Figure 1: Reality-Virtuality-Continuum [4]

According to Milgram et al. [4], the RV Continuum can be defined as *"… a generic Mixed Reality (MR) environment as one in which real-world and virtual world objects are presented together within a single display, that is, anywhere between the extrema of the RV continuum."*

In summary, AR can be understood as a way to enhance the users' perception and interaction with the real-world [3]. Although AR was firstly realized over 50 years ago, the use of this technology is still rudimental in both society and the economy. Identifying factors that influence the acceptance and further dissemination of AR is an important research question.

## B. Technology Acceptance

With emerging innovations, user acceptance refers to the decision whether to use a new technology or not. Acceptance is a latent construct with different factors influencing the construct. To explain these factors, different research models were developed for a wide variety of systems and applications in various domains [5][6]. The two most influential models of technology acceptance are the Technology Acceptance Model (TAM) by Davis [7] and the Unified Theory of Acceptance and Use of Technology (UTAUT) by Venkatesh et al. [8]. Both models are described in the following.

### 1) Technology Acceptance Model

The TAM is the best known and most widely empirically validated model for explaining user acceptance towards technology [9]. The research model was developed by Davis as part of his doctoral dissertation [7] and can be traced back to the Theories of Reasoned Action (TRA) and Planned Behavior (TPB). TRA and TPB are seen as important theoretical foundations of the TAM [7][9][10]. The TAM, including its variables and relationships, is shown in Figure 2.

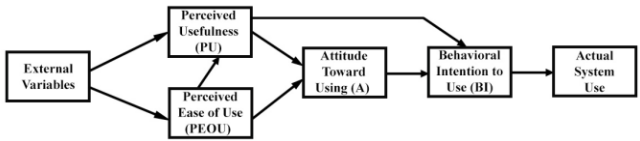

Figure 2: The Technology Acceptance Model (TAM) [10].

Davis assumed that the actual use of a system depends on the *Behavioral Intention (BI)* of the user. BI depends on the *User's Attitude towards Using (A)* and *Perceived Usefulness (PU)*. Together with *Perceived Ease of Use (PEOU)*, the two variables PU and PEOU build the core of the TAM: PU describes *"… the degree to which a person believes that using a particular system would enhance his or her job performance"*[7]. PEOU is defined as *"… the degree to which a person believes that using a particular system would be free of effort"*[7]. PU and PEOU have both a positive effect on the Attitude towards Using. In addition, PU and PEOU are influenced by *External Variables*. Furthermore, PU directly impacts BI [11].

Over the years, the TAM was frequently empirically validated and extended in different ways. Venkatesh et al. published two extended models in the form of the TAM 2 [12] and TAM 3 [13]. Since its introduction 35 years ago, the TAM has become a powerful and robust model for technology acceptance in the field of information systems research [6][9].

### 2) Unified Theory of Acceptance and Use of Technology

Another frequently used model in technology acceptance research [6] is the Unified Theory of Acceptance and Use of Technology (UTAUT) by Venkatesh et al. [8]. The UTAUT was developed by integrating the eight most common acceptance-related theories and their variables into a unified model. In the UTAUT, as presented in Figure 3, the three variables *Performance Expectancy*, *Effort Expectancy*, and *Social Influence*, directly influence the *Behavioral Intention*.

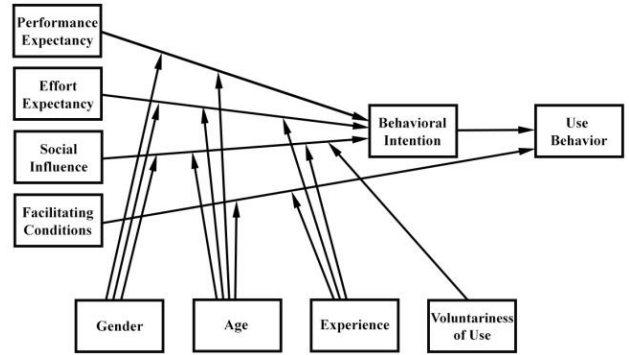

Figure 3: The Unified Theory of Acceptance and Use of Technology (UTAUT) [8].

*Performance Expectancy* is defined as *"… the degree to which an individual believes that using the system will help him or her to attain gains in job performance"* [8]. This construct is the strongest predictor of usage intention. Furthermore, *Effort Expectancy* is the second construct defined as *"… the degree of ease associated with the use of the system."* [8] The third construct is *Social Influence* which is introduced as *"… the degree to which an individual perceives that important others believe he or she should use the new system"*. The fourth construct, *Facilitating Conditions* represents *"… the degree to which an individual believes that an organizational and technical infrastructure exists to support use of the system"* [8]. *Facilitating Conditions* is the fourth main construct of the UTAUT and has a direct impact on use behavior. Additionally, the UTAUT model contains the variables *Gender*, *Age*, *Experience* and *Voluntariness of Use* as moderators, i.e., with moderating effects on the main variables [8].

## III. METHODOLOGY AND STUDY APPROACH

### A. Scope of the Study

Since their release, TAM and UTAUT have become the subject of many research papers. The models have been empirically validated in many different areas. In addition, other variables have been incorporated into the models to reflect the specifics of application environments or systems. The state of research and developments in technology acceptance research has already been the subject of many literature surveys [6][14]-[18]. The goal of all these studies was to provide a general overview of the models and the specific factors influencing technology acceptance. To the authors' knowledge, however, there is no work to date that more comprehensively examines the applied research models for technology acceptance in the field of AR and specifically its application in training and education. Our research aims to fill this research gap and provide a grounding for further analysis. The main contribution of this work is to gain an overview of the different models and TAM/UTAUT extensions, as well as variables applied in this field.

In our the literature review, we examined relevant research on AR technology acceptance according to the following criteria:
- *Research Objectives:* This refers to the aim of the research as well as to the AR systems, applications, and devices that have been investigated in the papers.
- *Sample Data*: Information about the geographic origin of the research and the user groups surveyed in the empirical part of the study.
- *Research Methods*: In addition, the evaluation method used for the models were analyzed.
- *Technology Acceptance Model*: The specific acceptance reseach model underlying the research.
- *Model Extension and Variables*: Extensions of the original models and corresponding acceptance variables.

*B. Systematic Literature Review Approach*

The literature review was based on a three-step approach, as shown in Figure 4. The objective was to gradually narrow down the articles to the point where research on technology acceptance of AR applications in training and education could be identified. The following seven scientific literature databases have been researched for relevant articles:

- *Google Scholar*
- *Science Direct*
- *Springer*
- *Emerald*
- *ACM Digital Library*
- *EconBiz*
- *IEEE Explore Digital Library*

In the first step, the terms *acceptance* and *augmented reality* were combined to form a search phrase. Only articles containing the search terms in their title were considered. This assumes that the title reflects the focus of the work. In summary, 204 articles were found.

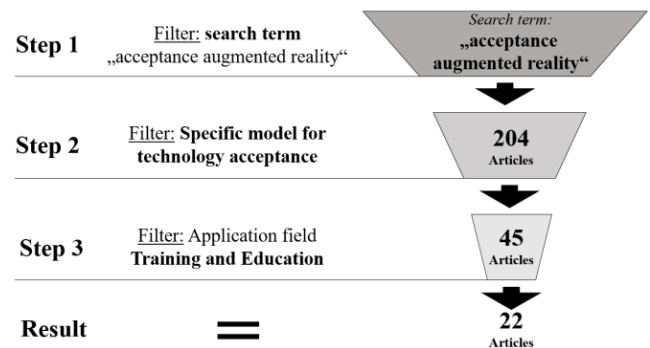

Figure 4: Approach to Literature Selection.

In the second step, all remaining articles were analyzed for referring to a specific technology acceptance model, i.e., the TAM or UTAUT and its variants, resulting in 45 articles.

In the third step, the specific application area of AR in training and application was applied as a filter criterion. As a result, 22 articles matching all criteria and focusing on technology acceptance of AR in training and education could be derived. The 22 articles are listed in Table III (Appendix).

IV. RESULT OF THE LITERATURE REVIEW

In the following, the results of the literature review are presented structured on the criteria as discussed in Section III.

*A. Research Objectives*

Ten research articles of the 22 identified have conducted a quantitative study, whereas only two used a qualitative approach. Ten research papers comprise mixed-method studies with quantitative as well as qualitative procedures. This result indicates that technology acceptance research is predominantly quantitative in nature.

Furthermore, the kind of AR device was examined. A distinction was made between mobile devices such as smartphones and tablets as well as AR glasses. Eight articles refer to mobile devices. No article investigated the use of AR glasses. Fourteen research papers did not specify a AR device as a subject of investigation at all.

*B. Sample Data*

Regarding the sample data, the countries of origin were first broken down by continent in which the surveys were conducted. Most contributions (ten papers) come from Asia, followed by seven articles from Europe. Three publications can be allocated to North America, whereas only one paper comes from South America. Two articles are cross-country studies that have conducted surveys in Australia and North America. This shows that most research activities are carried out in Asia and Europe.

*C. Research Methods*

The most popular research method for analyzing the sample data and validate the proposed research models are Structure Equation Modelling (SEM), regression analysis, and correlation analysis. These methods were used seven times in the 22 papers. Other evaluation methods, such as factor analysis, were applied only three times or less. The remaining papers did not provide any information on the evaluation methods.

*D. Technology Acceptance Model*

In this part, the different acceptance models were examined. Furthermore, the original acceptance model on which the articles are based was analyzed. In detail, 18 articles were based on the TAM. UTAUT and its variants built the basis for two articles. Furthermore, four papers did not propose an extension but applied the original models. The original TAM was used three times. The TAM2 was the referenced model in one research article. Another last article did not refer to any existing research model. The research activities show that the TAM is the most frequently used acceptance model, followed by the UTAUT and its successors.

These findings reflect the previous research activities in the field of technology acceptance research. As already men-tioned, especially the TAM is extended with different variables. The core components of the original TAM – *Perceived Usefulness (PU)* and *Perceived Ease of Use (PEOU)* have been integrated most. It must be declared that only the main constructs Attitude Toward Using (A), Behavioral Intention to Use (BI) and Actual System Use weren't considered. The third most used variable is *Perceived Enjoyment*. *Perceived Enjoyment* as introduced in TAM3 was defined as *"… the extent to which the activity of using the computer is perceived to be enjoyable in its own right, apart from any performance consequences that may be anticipated"* [13]. This variable was used in four of the AR acceptance research articles and thus most frequently for an extension of the TAM in all the papers investigated. Thus, the researchers seem to attach particular importance to this factor in the acceptance of AR. The other variables identified were used only very rarely. Against this background, the entire 45 identified AR acceptance research papers will be examined on the variables for a model extension in the following.

*E. Model Extensions and Variables*

In reference to the scope of this study, the different model extensions and variables are of importance. Thus, the last investigation criteria was applied to all 45 articles, which refer to the acceptance of AR and contain a specific technology acceptance model in all application fields. Of these 45 articles, 33 research articles have made an extension of a technology acceptance model by introducing own acceptance variables. Nineteen of these articles refer to training and education, 14 to other application fields. Figure 5 shows the composition of the article selection.

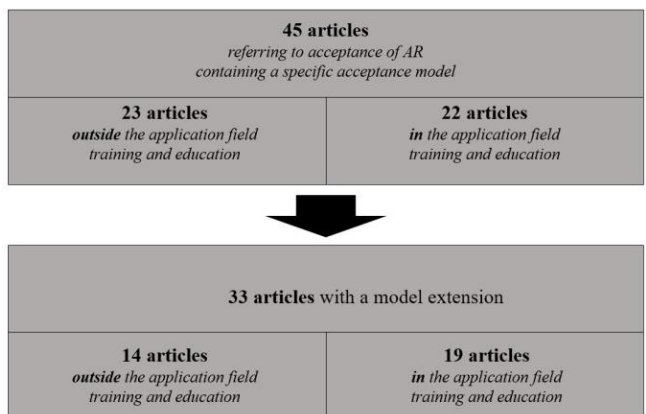

Figure 5: Composition of the article selection.

The following section examines the 14 articles that relate to AR applications in other fields before the model extension on training and education are discussed thereafter.

*1) AR Exensions Outside Training and Education*

Nine of the 14 articles outside training and education introduced variables on AR acceptance. The other five model extensions used already exisiting variables. In summary, 18 variables referring to the specific application area of AR were identified. The eight variables *Perceived Interactivity* [19], *Media Novelty* [19], *Previous Media Experience* [19], *Users' Innovativeness* [19], *Recommendation* [20], *Risk* [20], *Playfulness Expectancy* [21], and *Content Relevance Expectancy* [21] are introduced by the authors without providing a more comprehensive definition in the paper. The ten other variables are shown in Table I.

TABLE I. IDENTIFIED AR VARIABLES OUTSIDE TRAINING AND EDUCATION

| *Variable (Frequency of Use)* | *Definition [Source]* |
|---|---|
| Perceived Benefits/ Relative Benefit (2) | Positive aspects resulting from the use of AR [22][23][24] |
| Personal Innovativeness (2) | "Users' willingness to try out new services and products" [20][23][25] |
| Costs of Use (2) | Costs include efforts costs, loss of privacy costs and usage costs [20][23][26] |
| Self Presentation (1) | "Self-presentation is defined as presenting personal thoughts by using a creative manner of expression" [27] |
| Information Sharing (1) | "Information sharing refers to the level of willingness to share information with others" [27] |
| Visual appeal (1) | "Visual appeal relates to the exhibition of fonts and other visual elements such as graphics; it acts to enhance the overall presentation of information systems" [28] |
| Technology Readiness (1) | An overall state of mind that the user is ready to use a technology [28][29] |
| Personal Innovation (1) | "Users' willingness to adopt or reject a new technological innovation" [24] |
| Dimensions of cultural differences (1) | Different cultural dimensions which affect the technology acceptance. Uncertainty, Power distance, Masculinity-Feminity, Individualism/collectivism, and time orientation are summarized to the variable Dimensions of cultural differences [30]. |
| Personality Traits (1) | 'Big Five' Personality Factors conscienceless, Openness, Agreeableness, Neuroticism, and Extraversion [30]. |

The models are extended by the researchers with the aforementioned variables to propose AR acceptance models that are better aligned with the specific conditions of AR applications. The research model of Jung et al. [23] is shown here as an example.

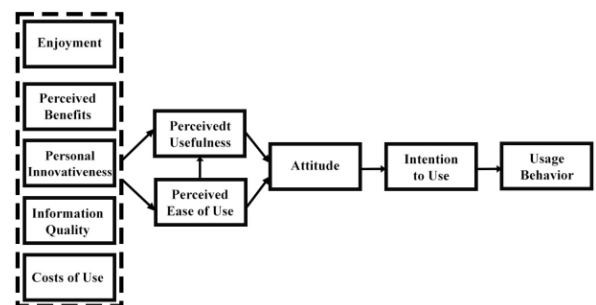

Figure 6: Proposed AR Acceptance Model [23].

The proposed model integrates variables from the TAM [7], IS Success Model [31], and UTAUT [8], as well as new own variables. This proposed research model refers to the application field tourism and urban heritage and represents one model extension as example. The example was chosen because of the mixture of three exisiting models as well as new variables. Furthermore, the base of the model can be allocated to the TAM as most of the model extensions investigated in this literature survey.

*2) AR Extensions in Training and Education*

As mentioned before, 19 papers with model extensions were identified for AR acceptance research in training and education. Twelve of these papers proposed research models with extended variables. In summary, 16 AR acceptance variables have been identified in the papers. Again, some papers introduced variables without a more comprehensive definition: *Teaching Experience* [32], *Technology Experiences* [32], *Characteristics of the system* [33], *Information Experience* [34], and *Information Literacy* [34]. One paper [35] proposed an acceptance model introducing *Duration of Use*, *Perceived Exertion*, *Emotion*, *Attachment*, *Harm*, *Perceived Change*, *Movement*, and *Anxiety* as *moderating factors* of AR acceptance. These six variables have not been further defined. In comparison, the other ten of the sixteen variables with their definition can be found in table II.

TABLE II. IDENTIFIED AR VARIABLES IN TRAINING AND EDUCATION

| Variable (Frequency of Use) | Definition [Source] |
|---|---|
| Perceived Situation Awareness (2) | Assistance for Understanding the environment around someone [36][37] |
| Interface Style (2) | Visualization of the AR content [36][37] |
| Technology Optimism (2) | "… a positive view of technology, including control, flexibility, convenience, and efficiency" [38][39] |
| Technology Innovativeness (2) | "… a person's inclination to try new information technologies" [38][39] |
| Visual Quality (1) | "… the degree to which a user considers that the app is aesthetically attractive to the eye" [40] |
| Ergonomics of AR-platform (1) | "The ergonomics of the ARTP refers to the features related to hardware and accessories that can help students develop favourable (or unfavourable) perceptions regarding the motivational factors." [41] |
| Resistance to Change (1) | "… attitudinal response of a person *not* accepting an innovation" [32] |
| Mobile Self-Efficacy (1) | "… an individual's perceptions of his or her ability to use mobile devices in order to accomplish particular tasks" [42][43] |
| Motivational Support (1) | "External support based on the culture, leadership and environment" [44] |
| Teachers' acceptance and integration of technology (TPACK) (1) | "… a theoretical framework which includes pedagogical knowledge, content knowledge, and pedagogical content knowledge for teaching. Furthermore, technology knowledge refers to these aspects." [44] |

The analysis shows that researchers in the field of training and education are also trying to adapt TAM and UTAUT to AR and the specific application conditions in the field through models with extended sets of variables. However, the low frequency of variable use also indicates that generally accepted extended models have not yet emerged. Rather, the models can be seen as individual attempts to adapt the traditional acceptance models to the scope of the particular study.

Koutromanos and Mikropoulos [42] developed a Mobile Augmented Reality Acceptance Model (MARAM) based on factors to explain teachers' intention to use educational AR applications.

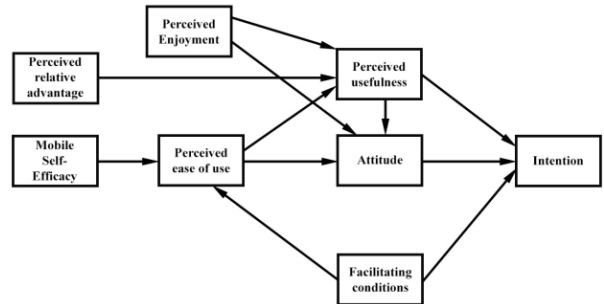

Figure 7: Mobile AR Acceptance Model (MARAM) [42].

In this proposed model, as shown in Figure 7, the acceptance variables introduced for model extension are *Perceived Relative Advantage (PRA)* and *Mobile Self-Efficacy*. PRA refers to the perception of the user whether the MAR Application is seen as better than conventional methods in education. Mobile Self-Efficacy regards the ability to use a mobile device. In addition, existing variables from the TAM variants, such as Perceived Enjoyment and Facilitating conditions, are applied next to the core variables of the TAM.

V. SUMMARY AND CONCLUSIONS

As a result of this literature review, it should first be noted that technology adaptation research in the field of AR is still a very young research field, with relatively few research papers published. Only 45 research articles with specified research models on technology acceptance of AR could be identified. Training and education was the most frequent area of application with 22 articles. With a close look at this field of application, it becomes clear that the papers primarily focus on the field of academic teaching. In comparison, educational applications outside schools and academics, e.g., using AR for training in companies or in industrial environments, show a lack of research.

Concerning the research method, it can be stated that in most cases, quantitative research (parts) were found. Regarding the regions of research, most research activities can be located in Asia as well as Europe. Looking at the research methods, it shows that Structure Equation Modelling (SEM), regression and correlation Analysis are the most commonly used methods for statistical analysis in acceptance research examined.

The research models and their extensions have been the focus of this study. The results show that the TAM and its core variables are the most frequently used models and theoretical foundation. Almost all research articles are based on the TAM or its extensions. In addition, the UTAUT and its extension can be stated ad well. Three-quarters of all investigated 45 papers on AR acceptance research with a specified model have made a model extension integrating own variables. In total, 34 AR variables not included in any exisiting acceptance model and its model variants have been identified. Despite the large number of variables, only a few relate specifically to the technology AR or the application field of training and education.

This study initially focused on the identification of the relevant papers on AR acceptance research and the variables for a model extension. The empirical significance of these variables as influencing factors and their importance in explaining AR acceptance need to be addressed in further research. However, an analysis is only possible for those studies that disclose relevant information on the corresponding statistical analysis.

It can be seen as problematic that researchers seem to focus on producing new models with their own and unique extended acceptance variables instead of contributing to the empirical validation of existing ones. Many of the specific variables found in our analysis are only applied by individual researchers. Thus, there is no generalizable AR technology acceptance model that has yet been sufficiently empirically validated in different application areas. Existing specialized AR research models, such as MARAM, should be validated more comprehensively for application in other contexts. Moreover, there is a particular lack of models that reflect application conditions in the field of corporate training outside schools and academic institutions.

Appendix

TABLE III. IDENTIFIED AR ACCEPTANCE RESEARCH PAPERS ON TRAINING AND EDUCATION

| Author (Year) | [Source] |
|---|---|
| A. Balog and C. Pribeanu (2010) | [41] |
| Y. Wang, A. Anne, and T. Ropp (2016) | [45] |
| C.-C. Mao, C.-C. Sun, and C.-H., Chen (2017) | [37] |
| C. Papakostas, C. Troussas, A. Krouska, and C. Sgouropoulou (2020) | [46] |
| T. Arvanitis, D. Williams, J. Knight, C. Baber, M. Gargalakos, S. Sotiriou and FX. Bogner (2020) | [35] |
| A. Hamed, K. Manolya and U.S. Nazim (2018) | [40] |
| J. Iqbal and M. S. Sidhu (2021) | [47] |
| A. Álvarez-Marín, J. A. Velázquez-Iturbide, and M. Castillo-Vergara (2021) | [38] |
| E. P. A. Sugara and Mustika (2016) | [48] |
| L. Ping and K. Liu (2020) | [49] |
| T. M. M. Alroqi (2021) | [50] |
| J. Iqbal and M. S. Sidhu (2019) | [51] |
| J. Ma, Q. Liu, S. Yu, M. Liu, J. Liu, and L. Wu (2021) | [33] |
| M. Al-Ani and N. Kasto (2018) | [52] |
| X. Geng and M. Yamada (2021) | [53] |
| A. Álvarez-Marín, J. A. Velázquez-Iturbide, and M. Castillo-Vergara (2021) | [54] |
| M. Ibáñez-Espiga, A. Di Serio, D. Villarán-Molina, and C. D. Kloos (2016) | [55] |
| J. Jang, Y. Ko, W. S. Shin, and A. I. Han (2021) | [44] |
| G. Banerjee and S. Walunj (2019) | [56] |
| J.-H. Lo and Y.-F. Lai (2018) | [34] |
| I. Vrellis, M. Delimitros, P. Chalki, P. Gaintatzis, I. Bellou, and T. A. Mikropoulos (2020) | [36] |
| G. Koutromanos and T. A. Mikropoulos (2021) | [42] |